\documentstyle[eqsecnum,aps,preprint]{revtex}
\newcommand{\st}{\stackrel}
\newcommand{\lh}{\leftrightarrow}
\newcommand{\pp}{\partial}
\newcommand{\s}{\sigma}
\newcommand{\p}{\perp}
\newcommand{\be}{\begin{eqnarray}}
\newcommand{\e}{\end{eqnarray}}
\begin{document}
\tighten
\title{ Transverse Spin in QCD. I. Canonical Structure}
\author{{\bf A. Harindranath}\thanks{e-mail: hari@tnp.saha.ernet.in}, 
{\bf Asmita Mukherjee}\thanks{e-mail: asmita@tnp.saha.ernet.in} \\
Saha Institute of Nuclear Physics, 1/AF, Bidhan Nagar, 
	Calcutta 700064 India \\
{\bf Raghunath Ratabole}\thanks{e-mail: raghu@cts.iisc.ernet.in} \\
 Centre for Theoretical Studies, Indian Institute of Science \\
     Bangalore 560012 India \\}
\date{April 25, 2000}
\maketitle
\begin{abstract}
In this work we initiate a systematic investigation of the spin of a
composite system in an arbitrary reference frame in QCD. After a brief review of the 
difficulties one encounters in equal-time quantization, we turn to 
light-front quantization. We show that, in spite of the complexities,
light-front field theory offers a unique opportunity to address the 
issue of relativistic spin operators in an arbitrary reference frame 
since boost is kinematical in this formulation. Utilizing this symmetry, 
we show how to introduce transverse spin operators for massless particles 
in an arbitrary reference frame in analogy with those for massive particles. 
Starting from the manifestly gauge invariant, symmetric energy momentum
tensor in QCD, we derive expressions for the interaction dependent 
transverse spin operators ${\cal J}^i$ ($i=1,2$) which are responsible 
for the helicity flip of the nucleon in light-front quantization. In order 
to construct ${\cal J}^i$, first we derive expressions for the transverse 
rotation operators $F^i$. In the gauge  $A^+=0$, we eliminate the 
constrained variables. In the completely gauge fixed sector, in terms of 
the dynamical variables, we show that one can decompose
${\cal J}^i= {\cal J}^i_I + {\cal J}^i_{II} + {\cal J}^i_{III}$ where only 
${\cal J}^i_{I}$ has explicit coordinate ($x^-, x^i$) dependence in its 
integrand. The operators ${\cal J}^i_{II}$ and ${\cal J}^i_{III}$ arise from 
the fermionic and bosonic parts respectively of the gauge invariant energy 
momentum tensor. We discuss the implications of our results.

\end{abstract}
\vskip .2in 
\section{Introduction}

From the early days of quantum field theory, it has been recognized that the
issues associated with the spin of a composite system in an arbitrary
reference frame are
highly complex and non-trivial\cite{alfaro}.
The familiar Pauli-Lubanski operators readily qualify for spin operators 
{\it only} in the rest frame of
the particle. For a single particle in a moving frame it is known\cite{gur} 
how to construct the appropriate spin operators starting from the 
Pauli-Lubanski operators. 
How to construct the spin operators for a composite system in an arbitrary
reference frame is a nontrivial problem. In equal-time quantization,  
the complexities arise
from the facts that for a moving composite object, {\it Pauli-Lubanski
operators are necessarily interaction dependent} and, further, it is quite
difficult\cite{os} to separate the center of mass and internal variables which is
mandatory in the calculation of spin. Due to these difficulties there has been
rarely any attempt to study the spin of a moving composite system in the
conventional equal time formulation of even simple field theoretic models,
let alone Quantum Chromo Dynamics (QCD).

From the early days of light-front field theory, the complications
associated with
transverse rotation operators $F^i$ 
have been recognized. They are interaction
dependent just like the Hamiltonian. Furthermore, together with the third
component of the rotation operator $J^3$, which is kinematical, $F^i$  do 
not obey the angular momentum algebra. Instead they obey the algebra of 
two dimensional Euclidean group which is appropriate only for massless
particles. For massive particles, one can define transverse spin operators
\cite{ls78} which together with the third component (helicity) obey the angular
momentum algebra. However, they cannot be separated into orbital and spin
parts unlike the helicity operator\cite{hk}. 
Most of the studies of the transverse spin operators in light-front field
theory, so far, are
restricted to free field theory\cite{except}. 
Even in this case the operators have a
complicated structure. However, one can 
write these operators as a
sum of orbital and spin parts, which can be achieved via a unitary
transformation, the famous Melosh transformation\cite{melosh}. 
In interacting
theory, presumably this can be achieved order by order\cite{bp} in a suitable
expansion parameter 
which is justifiable only in a weakly  coupled theory.

Knowledge about 
transverse rotation operators and transverse spin operators is mandatory 
for addressing issues concerning Lorentz invariance in
light-front theory. Unfortunately,
very little is
known\cite{review} regarding the field theoretic aspects of the
interaction dependent spin operators,  {\it We emphasize that in a moving 
frame, the spin
operators are interaction dependent irrespective of whether one considers
equal-time field theory or light-front field theory}. 
To the best of our knowledge, in gauge field theory, the canonical 
structure of spin
operators of a composite system in a moving frame  has 
never been studied.
In this work we initiate a systematic investigation of the spin of a
composite system in a moving frame
in QCD. A brief summary of some of our results has been presented in Ref.
\cite{lett}. 
We show that, in spite of the complexities,
light-front field theory
offers a unique opportunity to address the issue of relativistic spin
operators in an arbitrary reference frame since boost is kinematical in this
formulation.

The plan of this paper is as follows. In Sec. II, first, we briefly
review the complexities associated with the description of the spin of a
composite system in a moving frame in the conventional equal time
quantization. Then we give
the canonical
structure of light-front Lorentz algebra and light-front spin operators. 
In this section we also provide a detailed discussion of the transverse spin
operators for a massless particle of arbitrary transverse momentum.
The
explicit form of transverse spin operators in light-front QCD is derived
in Sec. III. 
Summary and conclusions are presented in Sec. IV. 
For the sake of completeness and clarity,
in Appendix A we review the intrinsic
spin operators in relativistic quantum mechanics.
The explicit form of the kinematical operators and the Hamiltonian in
light-front QCD starting from the gauge invariant, symmetric, interaction
dependent,  energy momentum
tensor is derived in Appendix B. A complete discussion of 
transverse spin operators in free fermion
field theory and free massless, spin one boson field theory 
is presented in detail in Appendices C and D.     

\section{Preliminaries}
In this section, first we briefly review the intrinsic spin operator in
equal-time quantization. We highlight the difficulties one encounters in
constructing the spin operator of a composite system in an arbitrary
reference frame in this case.
Next, we give the Lorentz generators in light-front
formulation and show that with the help of the kinematical boost in the
light-front formalism, a relativistic spin operator for a composite system
can be defined in an arbitrary reference frame for massive as well as
massless particles. We also compare and contrast
the spin operators in equal-time and light-front quantization.
\subsection{Intrinsic Spin in Equal Time Quantization}
Intrinsic spin operators in an arbitrary reference frame in equal-time 
quantization are given\cite{gur} in 
terms of the Poincare generators by (see Appendix
A for details)
\begin{eqnarray}
{\bf S}=&&{1\over M}\left[{\bf W}-{{\bf P}W^{0}\over {M+H}}\right]\nonumber\\
  &&={\bf J}~{P^0 \over M} - {\bf K} \times  {{\bf P}\over M} -
{({\bf J} \cdot {\bf P})\over {M+P^0}}{{\bf P}\over M}    
\end{eqnarray} 
where ${\bf W}$ are the space components of the Pauli-Lubanski
operator, $W^\mu=-{1\over 2}\epsilon^{\mu \nu \rho \lambda}M_{\nu \rho}
P_\lambda$. $ H $,${\vec P}$ are equal time Hamiltonian and momentum
operators respectively obtained by integrating the energy
momentum tensor over a spacelike surface and $ {\vec J}$ and $ {\vec K}$ are
the equal time rotation and boost generators respectively, which are obtained by
integrating the angular momentum density over a spacelike surface.  
Since boost ${\bf K}$ is dynamical, {\it all the three components of ${\bf S}$
are interaction dependent} in the equal time quantization. 
Nevertheless, the component of {\bf S} along {\bf P} remains kinematical.
This is to be
compared with light-front quantization where {\it the third component of the
light-front spin operator ${\cal J}^3$ is kinematical} (see Sec. IIB).
This arises from the facts that boost operators are kinematical on the light
front, the interaction dependence of light-front spin
operators ${\cal J}^i$ arises solely from the rotation operators, and the
third component of the rotation operator $J^3$ is kinematical on the light
front.

A further essential complication arises in equal time quantization. In order
to describe the intrinsic spin of a composite system, one should be able to
separate the center of mass motion from the internal motion. Even in free field
theory, this turns out to be quite involved (See Ref. \cite{os} and references
therein). On the other hand, in light-front theory, since transverse boosts
are simply Galilean boosts, separation of center of mass motion and internal
motion is as simple as in non-relativistic theory. (See Appendix A, 
of Ref. \cite{wip} for a detailed example).     
 
\subsection{Light-Front Lorentz Generators and Algebra}
In terms of the gauge invariant, symmetric energy momentum tensor
$\Theta^{\mu \nu}$, the four-vector $P^\mu$ and the 
tensor $M^{\mu \nu}$ are given by
\begin{eqnarray}
P^\mu &&= {1 \over 2} \int dx^- d^2 x^\perp \Theta^{+ \mu}, \\
M^{\mu \nu} && = {1 \over 2} \int dx^- d^2 x^\perp \left [ x^\mu \Theta^{+
\nu} - x^\nu \Theta^{+ \mu} \right ]. \label{def1}
\end{eqnarray}
The boost operators are $ M^{+-} = 2 K^3$ and $M^{+i}=E^i$. The rotation
operators are $ M^{12}=J^3$ and $ M^{-i} = F^i$. The Hamiltonian $P^-$ and
the transverse rotation operators $F^i$ are dynamical (depend on the
interaction) while other seven operators are kinematical. The 
rotation operators obey the $E(2)$-like algebra of two dimensional Euclidean 
group, namely,
\begin{eqnarray}
[F^1,F^2]=0, ~ [J^3,F^i] = i \epsilon^{ij} F^j
\end{eqnarray}
where $\epsilon^{ij}$ is the two-dimensional antisymmetric tensor. Thus $F^i$
do not qualify as angular momentum operators. Moreover, $F^i$ are not
translationally invariant and hence they do not qualify as intrinsic spin.  
\subsection{Transverse Spin Operators: Massive particle}
The Pauli-Lubanski spin operator 
\begin{eqnarray}
W^\mu = - { 1 \over 2} \epsilon^{\mu \nu \rho \sigma} M_{\nu \rho}
P_\sigma
\end{eqnarray}
with $ \epsilon^{+-12} = -2$.
For a massive particle, the transverse spin operators\cite{ls78} ${\cal J}^i$ in 
light-front theory are given in terms of Poincare generators by
\begin{eqnarray}
M{\cal J}^1 &&= W^1 - P^1 {\cal J}^3 = { 1 \over 2} F^2 P^+ + K^3 P^2   - { 1
\over 2} E^2 P^- - P^1 {\cal J}^3, \label{j1}
\e
\be  
M{\cal J}^2 &&= W^2 - P^2 {\cal J}^3 = - { 1 \over 2 } F^1 P^+ -K^3 P^1  + { 1 \over 2} E^1 P^- -
P^2 {\cal J}^3. \label{j2}
\end{eqnarray}
The first term in Eqs. (\ref{j1}) and (\ref{j2}) contains both center of
mass motion and internal motion and the next three terms in these equations
serve to remove the center of mass motion.  
 
The helicity operator is given by
\begin{eqnarray}
{\cal J}^3 &&= {W^+ \over P^+} = J^3 + { 1 \over P^+}(E^1P^2 - E^2 P^1).
\end{eqnarray} 
Here, $J^3$ contain both center of mass motion and internal motion and the
other two terms serve to remove the center of mass motion. 
The operators ${\cal J}^i$  obey the angular momentum commutation relations 
\begin{eqnarray}
\left [ {\cal J}^i, {\cal J}^j \right ] = i \epsilon^{ijk} {\cal J}^k
\label{j3} .
\end{eqnarray}
They commute with $P^\mu$. 
\subsection{Transverse Spin Operators: Massless case}
Again, we start from the Pauli-Lubanski spin operator,
\begin{eqnarray}
W^\mu = - { 1 \over 2} \epsilon^{\mu \nu \rho \sigma} M_{\nu \rho } p_\sigma
.
\end{eqnarray}
For the light-like vector $ p^\mu$, usually the collinear choice is
made\cite{Tung,Weinberg}, namely, $p^+ \neq 0$, $
p^\perp=0$. Then we get, $ W^-=0$, $ W^+ = J^3 p^+$, $W^1 = { 1 \over 2} F^2
p^+$, $ W^2 = - { 1 \over 2} F^1 p^+$. 

In free field theory, we have explicitly
constructed the Poincare generators for a massless spin one particle in
$A^+=0$ gauge in Appendix D. 
Consider the single particle state $ \mid p \lambda \rangle$ with $
p^\perp=0$. 
From the explicit form of the operators, we find that 
\begin{eqnarray}
J^3 \mid p \lambda \rangle &&= \lambda \mid p \lambda \rangle, \nonumber \\
F^i \mid p \lambda \rangle && = 0, ~ i=1,2 
\end{eqnarray}
since $ p^\perp=0$.

For calculations with composite states (dressed partons, for example) we
have to consider light-like particles with arbitrary
transverse momenta. Let us try a light like momentum $ P^\mu$ with $ P^\perp
\neq 0$, but $ P^- = {(P^\perp)^2 \over P^+}$ so that $P^2 = 0$. 
Then we get, as in the case of massive particle,
\begin{eqnarray}
W^+ && = J^3 P^+ + E^1 P^2 - E^2 P^1 ,\nonumber \\
W^1 && = { 1 \over 2} F^2 P^+ +K^3 P^2     - { 1 \over 2} E^2 P^-,\nonumber \\
W^2 && = - { 1 \over 2} F^1 P^+ - K^3 P^1  + { 1 \over 2} E^1 P^-, \nonumber
\\
W^- && = F^2 P^1 - F^1 P^2 - J^3 P^-.    
\end{eqnarray}
Thus even though $W^1$ and $W^2$ do not annihilate the state, we do get
$W^\mu W_\mu(={1\over 2}W^+W^- + {1\over 2}W^-W^+ - (W^1)^2 - (W^2)^2)
  \mid k \lambda \rangle =0$ 
as it should be for a massless particle.

Just as in the case of massive particle, we have the helicity operator for
the massless particle,
\begin{eqnarray}
{\cal J}^3 &&= {W^+ \over P^+} = J^3 + { 1 \over P^+}(E^1P^2 - E^2 P^1).
\end{eqnarray}
In analogy with 
the transverse spin for massive particles, we define the transverse spin 
operators for massless particles as
\begin{eqnarray}
{\cal J}^i = W^i - P^i {\cal J}^3.
\end{eqnarray}
They do satisfy
\begin{eqnarray}
{\cal J}^i \mid k \lambda \rangle &&=0, \nonumber \\
{\cal J}^3 \mid k \lambda \rangle && = \lambda \mid k \lambda \rangle, 
\end{eqnarray}
where $k$ is an arbitrary momentum. 
The operators ${\cal J}^i$ and ${\cal J}^3$ obey the $E(2)$-like algebra
\begin{eqnarray}
\left [{\cal J}^1, {\cal J}^2 \right ] =0, ~
\left [{\cal J}^3, {\cal J}^1 \right ] = i {\cal J}^2,~
\left [{\cal J}^3, {\cal J}^2 \right ] = -i {\cal J}^1.  
\end{eqnarray}
\subsection{Comments}
In order to calculate the transverse spin operators, first we need to
construct the Poincare generators $P^+$, $P^i$, $P^-$, 
$K^3$, $E^i$, $J^3$ and
$F^i$ in light-front QCD. The explicit form of the operator $J^3$ is given
Ref. \cite{hk}. The construction of $F^i$ which is algebraically 
quite involved is carried out in the next section. The construction of the
rest of the kinematical operators is given in Appendix B. In this appendix we also present
the Hamiltonian in a manifestly Hermitian form.    

In order to have a physical picture of the complicated situation at hand it
is instructive to calculate the spin operator in free 
field
theory. The case of free massive fermion is carried out in Appendix C. 
In free field theory one can
indeed show that (see Appendix C) ${\cal J}^i \mid k \lambda \rangle = 
{ 1 \over 2} \sum_{\lambda'} \sigma^i_{\lambda' \lambda} \mid k \lambda'
\rangle$.   The case of free massless spin one
particle is carried out in Appendix D.
\section{The transverse rotation operator in QCD}
In this section we derive the expressions for interaction dependent
transverse rotation operators in light-front QCD starting from the
manifestly gauge invariant energy momentum tensor. It is extremely
interesting to compare and contrast the situation in the equal time and
light-front case. 
The angular momentum density
\begin{eqnarray}
{\cal M}^{\alpha \mu \nu} = x^\mu \Theta^{\alpha \nu} - x^\nu
\Theta^{\alpha \mu}. 
\end{eqnarray}
In equal time theory, generalized angular momentum
\begin{eqnarray}
M^{\mu \nu} = \int d^3x {\cal M}^{0 \mu \nu}.
\end{eqnarray}
The rotation operators are $ J^i = \epsilon^{ijk} M^{jk}$. Thus in a
non-gauge theory, all the three components of the rotation operators are
manifestly interaction independent. However, the spin operators $S^i$ for a 
composite system in
a moving frame involves, in addition to  $J^i$, the  boost operators
$K^i = M^{0i}$ which are interaction dependent. {\it Thus all the three
components of $S^i$ become interaction dependent.}

A gauge invariant separation of the nucleon angular momentum is performed in
Ref. \cite{ji}. However, as far the spin operator in an arbitrary reference
frame is concerned, 
the analysis of this reference is valid only in the rest frame where spin
coincides with total angular momentum operator and in an arbitrary 
reference frame the need to project out the
center of mass motion, which is quite complicated in equal time theory is
not emphasized. 
Moreover, the distinction between the longitudinal and transverse components
of the spin is never made.
It is crucial to make this distinction since physically
the longitudinal and transverse components of the spin carry quite distinct
information (as is clear, for example, from the spin of a massless particle). 
Moreover, even for the third component of the spin of a composite system
in a moving frame, there is crucial difference between equal time and light
front cases. ${\cal J}^3$
(helicity) is interaction independent whereas $S^3$ is interaction
dependent in general except when measured along the direction of {\bf P}.

In light-front theory, generalized angular momentum 
\begin{eqnarray}
M^{\mu \nu} =
{ 1 \over 2} \int dx^- d^2 x^\perp {\cal M}^{+ \mu \nu}.
\end{eqnarray}
$J^3$ which is related to the helicity is given by  
\begin{eqnarray}
J^3 = M^{12} = { 1
\over 2} \int dx^1 d^2 x^\perp [x^1
\Theta^{+2} - x^2 \Theta^{+1} ] 
\end{eqnarray}
and is interaction independent.
On the other hand, the transverse rotation operators which are related
to the transverse spin are given by
$$ F^i =M^{-i}= { 1 \over 2} \int dx^- d^2 x^\perp [ x^- \Theta^{+i} - x^i
\Theta^{+-} ] . $$
They are interaction dependent {even in a non-gauge theory} since
$\Theta^{+-}$ is the Hamiltonian density.

In light-front theory we set the gauge $A^+=0$
and eliminate the dependent variables $\psi^-$ and $A^-$ using the equations
of constraint. In this paper we restrict to the topologically trivial sector
of the theory and set the boundary condition $A^i(x^-, x^i) \rightarrow 0 $
as $ x^{-,i} \rightarrow \infty$. This completely fixes the gauge and put
all surface terms to zero.

The transverse rotation operator 
\begin{eqnarray}
F^i = {1 \over 2} \int dx^- d^2 x^\perp \Big [ x^- \Theta^{+i} - x^i
\Theta^{+-} \Big ].
\end{eqnarray}
The symmetric, gauge invariant energy momentum tensor 
\begin{eqnarray}
\Theta^{\mu \nu} &&= { 1 \over 2} {\overline \psi} \Big [ 
  \gamma^\mu i D^\nu + \gamma^\nu i D^\mu \Big ] \psi - F^{\mu \lambda a}
F^{\nu a}_{\, \, \lambda} 
 - g^{\mu \nu} \Big [ - { 1 \over 4} (F_{\lambda \sigma a})^2 +
{\overline \psi} ( \gamma^\lambda i D_\lambda - m) \psi \Big ],
\end{eqnarray}
where 
\begin{eqnarray}
i D^\mu &&= {1 \over 2} \st{\lh}{i\pp^\mu} + g A^\mu, \nonumber \\
F^{\mu \lambda a} && = \partial^\mu A^{\lambda a} - \partial^\lambda A^{\mu
a} + g f^{abc} A^{\mu b} A^{\lambda c}, \nonumber \\
F^{\nu a}_{\, \, \lambda} && = \partial^\nu A_{\lambda}^a - \partial_\lambda
A^{\nu a} + g f^{abc} A^{\nu b} A_\lambda^c.
\end{eqnarray}
First consider the fermionic part of $ \Theta^{\mu \nu}$:
\begin{eqnarray}
\Theta^{\mu \nu}_F = { 1 \over 2} {\overline \psi} \Big [ \gamma^\mu i D^\nu
+ \gamma^\nu i D^\mu \Big ]\psi - g^{\mu \nu } {\overline \psi} (\gamma^\lambda
i D_\lambda - m)\psi.
\end{eqnarray}
The coefficient of $g^{\mu \nu}$ vanishes because of the equation of motion. 

Explicitly, the contribution to $F^2$ from the fermionic part of
$\Theta^{\mu \nu}$ is given by
\begin{eqnarray}
F^2_F && = { 1 \over 2} \int dx^- d^2 x^\perp \left [ x^- { 1 \over 2}
{\overline \psi} (\gamma^+ i D^2 + \gamma^2 i D^+) \psi 
 - x^2 { 1 \over 2}{\overline \psi} (\gamma^+ i D^- + \gamma^- i D^+) \psi
\right ],  \nonumber \\
&& = F^2_{F(I)} + F^2_{F(II)},
\end{eqnarray} 
where 
\begin{eqnarray}
F^2_{F(I)}= { 1 \over 2} \int dx^- d^2 x^\perp x^- \Big [ 
{\psi^+}^\dagger {1 \over 2} \st{\lh}{i\pp^2} \psi^+ + {\psi^+}^\dagger g A^2 \psi^+ + 
{ 1 \over 4} {\overline
\psi} \gamma^i  \st{\lh}{i\pp^+}
\psi \Big ],
\end{eqnarray}
\begin{eqnarray}
F^2_{F(II)}= -{ 1 \over 2} \int dx^- d^2 x^\perp x^2 \Big [ 
{\psi^+}^\dagger \Big ({1 \over 2} \st{\lh}{i\pp^-} + gA^- \Big ) \psi^+ 
 + 
{ 1 \over 4}
{\psi^-}^\dagger \gamma^i \st{\lh}{i\pp^+}
\psi^- \Big ].
\end{eqnarray} 
We have the equation of constraint
\begin{eqnarray}
i \partial^+ \psi^- = \big [ \alpha^\perp \cdot ( i \partial^\perp + g
A^\perp) + \gamma^0 m \big ] \psi^+, \label{eoc}
\end{eqnarray}
and the equation of motion
\begin{eqnarray}
i \partial^- \psi^+ = -g A^- \psi^+ + \big [ \alpha^\perp \cdot  (i
\partial^\perp + g A^\perp) + \gamma^0 m \big]{ 1 \over i \partial^+}
\big [ \alpha^\perp \cdot  (i
\partial^\perp + g A^\perp) + \gamma^0 m \big]   \psi^+. \label{eom}
\end{eqnarray}
Using the Eqs. (\ref{eoc}) and (\ref{eom}) we arrive at free ($g$
independent) and
interaction ($g$ dependent) parts of $F^2_F$.
The free part of $F^2_F$ is given by
\begin{eqnarray}
F^2_{F(free)} &&= { 1 \over 2} \int dx^- d^2 x^\perp \Bigg \{x^- \Bigg [
\xi^\dagger \Big [i \partial^2 \xi\Big] - 
\Big [i \partial^2 \xi^\dagger\Big ] \xi \Bigg ]  \nonumber \\
&&~~~~~~ - x^2 \Bigg [ \xi^\dagger \Big [{ - (\partial^\perp)^2 +m^2 \over i
\partial^+} \xi \Big ] -  \Big [{ - (\partial^\perp)^2 +m^2 \over i
\partial^+} \xi^\dagger\Big ] \xi \Bigg ] \nonumber \\
&& ~~~~~~+ \Bigg [ \xi^\dagger \Big [ \sigma^3 \partial^1 + i \partial^2
\Big]{ 1 \over
\partial^+} \xi + \Big [ { 1 \over \partial^+} (\partial^1 \xi^\dagger \sigma^3 -
i \partial^2 \xi^\dagger) \Big ] \xi \Bigg ] \nonumber \\ 
&&~~~~~~ + m \Bigg [ \xi^\dagger \Big [{ \sigma^1 \over i \partial^+} 
\xi\Big ] -
\Big [{ 1 \over i \partial^+} \xi^\dagger\sigma^1\Big ] \xi \Bigg ]
\Bigg \}.
\end{eqnarray}
We have introduced the two-component field $\xi$, 
\begin{eqnarray} \psi^+ = 
\left [ \begin{array}{c} \xi \\
                       0 \end{array} \right ].
\end{eqnarray} 
The interaction dependent part of $F^2_{F(I)}$ is 
\begin{eqnarray}
F^2_{F(I)int} && = g \int dx^- d^2 x^\perp x^- \xi^\dagger A^2 \xi \nonumber
\\
&& ~~~~+ { 1 \over 4} g \int dx^- d^2 x^\perp \Big [ \xi^\dagger { 1 \over
\partial^+}[(-i \sigma^3 A^1 + A^2)\xi] + { 1 \over \partial^+}
[ \xi^\dagger (i \sigma^3 A^1 + A^2)]\xi \Big ].
\end{eqnarray} 

The interaction dependent part of $F^2_{F(II)}$ is 
\begin{eqnarray}
F^2_{F(II)int} =  { 1 \over 4} g \int dx^- d^2 x^\perp \Big 
[ \xi^\dagger { 1 \over
\partial^+}[(-i \sigma^3 A^1 + A^2)\xi] + { 1 \over \partial^+}
[ \xi^\dagger (i \sigma^3 A^1 + A^2)]\xi \Big ] \nonumber \\
- { 1 \over 2} g \int dx^- d^2 x^\perp x^2 \Bigg [
{\pp^\p\over {\pp^+}} [ \xi^\dagger ({\tilde \s}^\p \cdot A^\p) ] {\tilde \s}^\p
\xi + \xi^\dagger ({\tilde \s}^\p \cdot A^\p) {1\over {\pp^+}} ({\tilde
\s}^\p \cdot \pp^\p) \xi \nonumber\\ ~~~~~ + ({ \pp^\p \over {\pp^+}}
\xi^\dagger) {\tilde \s}^\p ( { \tilde \s}^\p \cdot A^\p) \xi + \xi^\dagger
{1\over {\pp^+}} ({\tilde \s}^\p \cdot  \pp^\p) ( {\tilde \s}^\p \cdot
A^\p) \xi \nonumber\\ ~~~~~~ -m {1\over {\pp^+}} [ \xi^\dagger ({\tilde
\s}^\p \cdot A^\p) ] \xi + m \xi^\dagger ( {\tilde \s}^\p \cdot A^\p){1\over
{\pp^+}} \xi
\nonumber\\~~~~~~~~~~ + m ( {1\over {\pp^+}} \xi^\dagger) ({\tilde \s}^\p
\cdot A^\p) \xi - m \xi^\dagger {1\over {\pp^+}} [( {\tilde \s}^\p \cdot
A^\p) \xi] \Bigg ] \nonumber\\  
~~~~- {1 \over 2 } g^2 \int dx^- d^2 x^\perp x^2 \Bigg [ 
\xi^\dagger {\tilde \sigma}^\perp \cdot A^\perp  { 1 \over i \partial^+} 
{\tilde \sigma}^\perp \cdot (A^\perp \xi)
- { 1 \over i \partial^+} (\xi^\dagger {\tilde \sigma}^\perp \cdot A^\perp) 
{\tilde \sigma}^\perp \cdot A^\perp \xi \Bigg ].
\end{eqnarray} 
We have introduced $ {\tilde \sigma}^1 =  \sigma^2$ and $ {\tilde
\sigma}^2 = - \sigma^1$.

Next consider the gluonic part of the operator $F^2$:
\begin{eqnarray}
F^2_g = { 1 \over 2} \int dx^- d^2 x^\perp \Big [ x^- \Theta^{+2}_g - x^2
\Theta^{+-}_g \Big ],
\end{eqnarray} 
where
\begin{eqnarray} 
\Theta^{+2}_g &&= - F^{+ \lambda a} F^{2a}_{\, \, \lambda}, \nonumber \\
\Theta^{+-}_g &&= - F^{+\lambda a} F^{- a }_{\, \, \lambda} + { 1 \over 4}
g^{+-}(F_{\lambda \sigma a})^2.
\end{eqnarray}
Using the constraint equation
\begin{eqnarray}
{ 1 \over 2} \partial^+ A^{-a} = \partial^i A^{ia} + g f^{abc} { 1 \over
\partial^+}(A^{ib} \partial^+A^{ic}) + 2 g { 1 \over \partial^+} \Big (
\xi^\dagger T^a \xi \Big ),
\end{eqnarray}
we arrive at
\begin{eqnarray}
F^2_g = F^2_{g(free)} + F^2_{g(int)}
\end{eqnarray}
where
\begin{eqnarray}
F^2_{g(free)} &&= { 1 \over 2} \int dx^- d^2 x^\perp \Bigg \{ x^- \Big (
A^{ja}\partial^+\partial^j A^{2a} - A^{2a}\partial^+ \partial^j A^{ja}+
A^{ja}\partial^+\partial^2 A^{ja}\Big ) \nonumber \\
&& ~~~~~~~~~~~ -x^2  \Big ( A^{ka}(\partial^j)^2 A^{ka} \Big ) \Bigg
\}
\nonumber \\
&& ~~~~~~~~~~~~~~~~~~~- 2\int dx^- d^2 x^\perp A^{2a} \partial^1  A^{1a}.
\end{eqnarray}
The interaction part
\begin{eqnarray}
F^2_{g(int)} &&= { 1 \over 2} \int dx^- d^2 x^\perp x^- \Bigg \{ 
gf^{abc} \partial^+ A^{ia} A^{2b} A^{ic} \nonumber \\
&&~~~~ + g\Big (  f^{abc} { 1 \over \partial^+}(A^{ib} 
\partial^+ A^{ic}) + 2{
1 \over \partial^+} (\xi^\dagger T^a \xi) \Big ) \partial^+ A^{2a} \Bigg \}
\nonumber \\
&&~~ - {1 \over 2} \int dx^- d^2 x^\perp  x^2 \Bigg \{
2g f^{abc} \partial^i A^{ja} A^{ib} A^{jc} + {g^2 \over 2} f^{abc} f^{ade}
A^{ib} A^{jc} A^{id} A^{je} \nonumber \\
&& ~~~~ + 2g \partial^i A^{ia} { 1 \over \partial^+}
\Big ( f^{abc}  A^{jb} \partial^+ A^{jc} + 2 \xi^\dagger T^a \xi \Big )
\nonumber \\
&& ~~~~ + g^2 \Big ( f^{abc} { 1\over \partial^+} (A^{ib}\partial^{+} A^{ic})
+2 { 1 \over \partial^+} \xi^\dagger T^a \xi \Big )
\Big ( f^{ade} { 1\over \partial^+} (A^{jd}\partial^{+} A^{je})
+ 2{ 1 \over \partial^+} \xi^\dagger T^a \xi \Big ) \Bigg \}.  
\end{eqnarray}
So the full transverse rotation operator in QCD can be written as,
\begin{eqnarray}
F^2  = F^2_{I} + F^2_{II} + F^2_{III},
\end{eqnarray}
where
\begin{eqnarray}
F^2_{I}&& = {1\over 2} \int dx^- d^2x^\p [ x^- {\cal P}^2_0 - x^2 ({\cal H}_0 +
{\cal V}) ], \\
F^2_{II} &&= 
{1\over 2} \int dx^- d^2x^\p \Bigg [\xi^\dagger \Big [ \sigma^3 \partial^1 + i \partial^2
\Big]{ 1 \over
\partial^+} \xi + \Big [ { 1 \over \partial^+} (\partial^1 \xi^\dagger \sigma^3 -
i \partial^2 \xi^\dagger) \Big ] \xi \Bigg ] \nonumber \\ 
&&~~~~~~ + {1\over 2} \int dx^- d^2x^\p m \Bigg [ \xi^\dagger \Big [{ \sigma^1 \over i \partial^+} 
\xi\Big ] -
\Big [{ 1 \over i \partial^+} \xi^\dagger\sigma^1\Big ] \xi \Bigg ]
\nonumber \\
&& ~~+ {1\over 2} \int dx^- d^2x^\p  g \Bigg [ \xi^\dagger { 1 \over
\partial^+}[(-i \sigma^3 A^1 + A^2)\xi] + { 1 \over \partial^+}
[ \xi^\dagger (i \sigma^3 A^1 + A^2)]\xi \Bigg ], \\
F^2_{III}&&= 
- \int dx^- d^2 x^\perp 2 (\partial^1 A^{1})A^2 \nonumber \\
&&~~-{1\over 2} \int dx^- d^2x^\p g {4\over {\pp^+}} (\xi^\dagger T^a
\xi) A^{2a} - {1\over 2} \int dx^- d^2x^\p g f^{abc} {2\over {\pp^+}} (
A^{ib} \pp^+ A^{ic} ) A^{2a}
\end{eqnarray}
where $ {\cal P}^i_0$ is the free momentum density, $ {\cal H}_o$ is the
free Hamiltonian density and ${\cal V}$ are the interaction terms in the
Hamiltonian in manifestly Hermitian form (see Appendix B). The operators $F^2_{II}$ and $F^2_{III}$ 
whose integrands do not
explicitly depend upon coordinates arise from the fermionic and bosonic
parts respectively of the gauge invariant, symmetric, energy momentum tensor
in QCD. The above separation is slightly different from that in \cite{lett}. 
From Eq. (\ref{j1}) in Sec. II it follows that the transverse spin
operators ${\cal J}^i$, ($i=1,2$) can also be written as the sum of three
parts, ${\cal J}^i_{I}$ whose integrand 
has explicit coordinate dependence, ${\cal
J}^i_{II}$ which arises from the fermionic part, and  ${\cal J}^i_{III}$ which
arises from the bosonic part of the energy momentum tensor.   
\section{Summary and Conclusions}

We have initiated the study of spin operators in QCD. In equal
time quantization, one encounters two major difficulties in the description
of the spin of a composite system in an arbitrary reference frame. They are
1) the complicated interaction dependence arising from dynamical boost
operators and 2) the
difficulty in the separation of center of mass motion from the internal
motion. Due to these severe difficulties, there have been hardly any attempt
to study spin operators of a moving composite system in the conventional
equal time formulation of quantum field theory. 

In light-front theory, on the other hand, the longitudinal spin
operator (light-front helicity) is interaction independent and the
interaction dependence of transverse spin operators arises solely from that
of transverse rotation operators. Moreover, in this case the separation of
center of mass motion from internal motion is trivial since light-front
transverse boosts are simple Galilean boosts.    

We have investigated the case of transverse spin operators for both 
massive and massless particles. A novel feature here is the introduction 
of transverse spin operators for massless particles with arbitrary 
transverse momentum. {\it To the best of our knowledge, this is done for the
first time in light-front field theory}. 
To provide physical intuition for transverse spin 
operators which have a complicated structure in interaction theory, 
we have provided the explicit form of these operators in Fock space 
basis for both free fermion field theory and free massless spin 
one field theory. 

In QCD, our starting point is the formula for transverse rotation operators
expressed as the integral of generalized angular
momentum density given in terms of gauge invariant, symmetric, energy
momentum tensor. We have emphasized  the differences between spin operators
in field theory in equal time and light-front quantization schemes.  

Appropriate to light-front quantization, we choose the
light-front gauge. 
We use the constraint equations for $\psi^-$ and $A^-$ to eliminate them in
favor of dynamical degrees of freedom.  
In this initial study, we restrict to topologically
trivial sector of QCD and set the requirement that the transverse gauge
fields vanish as $x^{-,i} \rightarrow \infty$. This eliminates the
surface terms and completely fixes the gauge.
In the gauge fixed theory we found that the transverse rotation operators
can be decomposed as the sum of three distinct terms: $F^i_{I}$ which has
explicit coordinate dependence in its integrand, and $F^i_{II}$ and $F^i_{III}$ which have no
explicit coordinate dependence in their integrand. 
Further, $F^i_{II}$ and $F^i_{III}$ arise
from the fermionic and bosonic parts of the energy momentum tensor. 
Since transverse spin is responsible for the helicity flip of the nucleon in
light-front theory, we now have identified the complete set of
operators responsible for the helicity flip of the nucleon.

It is extremely interesting to contrast the 
cases of longitudinal and transverse spin operators
in light-front field theory. In the case of longitudinal spin operator
(light-front helicity), in the gauge fixed theory, the operator is
interaction independent and can be separated into orbital and spin parts for
quarks and gluons. It is known for a long time that the transverse spin 
operators in
light-front field theory cannot be separated into orbital and spin parts
except in the trivial case of free field theory.
{\it In this work, we have shown that, in spite of the complexities, 
a physically interesting separation is indeed
possible for the transverse spin operators} which is quite different from
the separation into orbital and spin parts in the rest frame familiar in the
equal time picture.    

In light-front theory, in addition to the
Hamiltonian, transverse spin operators  also contain interactions  and have a complicated structure. 
Since  transverse rotational symmetry is not manifest in light-front theory 
a study of these operators is essential for questions regarding Lorentz invariance
in the theory\cite{glazek}. 
An important issue in the case of transverse spin operators concerns
renormalization. Since they are interaction dependent, they will acquire
divergences in perturbation theory just like the Hamiltonian. It is of
interest to find the physical meaning of these divergences and their
renormalization. We address these issues in Ref. \cite{wip}
by computing the
expectation value of the transverse spin operators in a dressed quark state.

In this work we have explored in detail the theoretical aspects of spin 
operators in quantum field theory in the context of QCD and their
consequences.
Our construction and decomposition of the transverse spin operators in QCD also
have important phenomenological consequences. Elsewhere, we have 
shown\cite{lett} that nucleon expectation values of $F^i_{II}$ and
$F^i_{III}$ are directly related to the integrals of quark and gluon
distribution functions that appear in transversely polarized deep inelastic
scattering.
Our results show that one can relate nucleon expectation values
of operators appearing in the
transverse spin to transversely polarized deep inelastic scattering. It is
interesting to establish a transverse spin sum rule in analogy to the
helicity sum rule and explore its phenomenological consequences\cite{wip}.

\acknowledgments

We acknowledge helpful conversations with Rajen Kundu, Samir
Mallik, Partha Mitra, Jianwei Qiu and
James P. Vary. RR gratefully acknowledges the financial assistance of the
Council of Scientific and Industrial Research (CSIR), India.
\appendix
{\section{Intrinsic Spin in Relativistic Quantum Mechanics}
In this appendix, for the sake of clarity and completeness we review the
intrinsic spin operators in relativistic quantum mechanics \cite{jsc}.
The unitary representations of the Poincare group can be usefully classified on the
basis of sign of $M^2$, where $M^2=P^\mu P_\mu$
 ( and further by the sign of $H$ in case $M^2\geq 0$).
We consider two classes of representations which are of physical importance:
\begin{itemize}
{\item Positive time-like representations: $M^{2}>0~~H>0$}
{\item Positive light-like representations: $M^{2}=0~~H>0$}
\end{itemize}
In either cases we do not demand that the representations be irreducible (this
allows us to deal with elementary and composite systems simultaneously).
\subsubsection{ Positive time-like representations}
Beginning from the basic generators $H, {\bf P}, {\bf J}$, and ${\bf K}$ 
one can
construct an operator ${\bf S}$ such that it is translationally invariant, transforms
as a three vector under pure rotations and within itself obeys $SU(2)$ commutation
relations.
\be
[S^j,P^\mu]=0, ~~[J^j,S^k]=i \epsilon^{jkl}S^l, ~~[S^j,S^k]=i
\epsilon^{jkl}S^l.
\e
A suitable solution to the above requirements is provided by
\be
{\bf S}=&&{1\over M}\left[{\bf W}-{{\bf P}W^{0}\over {M+H}}\right]\nonumber\\
  &&={\bf J}~{P^0 \over M} - {\bf K} \times  {{\bf P}\over M} -
{({\bf J} \cdot {\bf P})\over {M+P^0}}{{\bf P}\over M}    
\e
where ${\bf W}$ are the space components of the Pauli-Lubanski
operator, $W^\mu=-{1\over 2}\epsilon^{\mu \nu \rho \lambda}M_{\nu \rho}
P_\lambda$.

The operators ${\bf S}$ cease to be defined when $M$ tends to zero.
The commutation relations among ${\bf P}, {\bf S}$ and $M$ are
given by  
\be
 [ P^j,S^k ]=0,~~
[S^j,S^k]=i \epsilon^{jkl}S^l,~~[S^j,M]=0.\label{rel1}
\e
Since ${\bf P}$ and $M$ stand for the momentum and invariant mass of the system,
the above relations make clear that ${\bf S}$  should represent
`intrinsic spin' of the system. 

The invariant $W^{2}$ can be completely expressed in terms of $M$ and ${\bf S}$ as
\be
W^{2}=-M^{2}{\bf S}^{2}.
\e
\vskip .2in
{\subsubsection{Positive light-like representations}

Begining from the basic generators ${\bf P}$, $\bf J$ and $\bf K$ (here
$H=|{\bf P}|$) one has to construct operators $S$, ${\cal T}^1$ and
${\cal T}^2$ such that they commute with four momentum $P^{\mu}$ and amongst 
themselves satisfy $E(2)$ commutation relations:
\begin{equation}
\begin{array}{lll}
[S,{\cal T}^1]=i{\cal T}^2, & [S,{\cal T}^2]=-i{\cal T}^1, & 
[{\cal T}^1,{\cal T}^2]=0.
\end{array} \end{equation}
A suitable solution consistent with the above requirements is:
\be
S={W^{0}\over \mid{\bf P}\mid }, \nonumber\\
{\cal T}^1=W^{1}-P^{1}{(W^{3}+W^{0})\over (\mid {\bf P}\mid+P^{3})},
\nonumber\\
{\cal T}^2=W^{2}-P^{2}{(W^{3}+W^{0})\over (\mid {\bf P}\mid +P^{3})}.
\e
Note that although ${\cal T}_1$ and ${\cal T}_2$ coincide with the front 
definitions, the difference lies in the remaining component. Note that here
$S$ is the component of angular momentum in the direction of motion. 
To further bring out the difference, we note in passing that $S$ is a scalar
under pure spatial rotation, while shows complicated behaviour under
pure boosts. Contrast this with the fact that ${\cal J}^3$ is front boost
invariant.
\subsubsection{Comments}
 The generators for a multi-particle relativistic 
system have been analyzed by several authors \cite{os}. The expressions
obtained are too complicated to be used in any practical calculations and
the generators cannot
be easily separated into the center of mass and internal variables.
Moreover, the derivations have been done neglecting the field theoretical
effects such as pair creation and crossing and so are expected to be valid
in the relatively low energy region where an expansion in ${v\over c}$ is
permissible. Interactions are to be incorporatated by introducing an
effective potential which vanish sufficiently rapidly for large distance.

\section{Poincare generators in light-front QCD}
In this appendix we derive the manifestly hermitian kinematical Poincare 
generators (except
$J^3$) and the Hamiltonian in light-front QCD
starting from the gauge invariant symmetric energy momentum tensor
$\Theta^{\mu \nu}$. To begin with, $\Theta^{\mu \nu}$ is interaction
dependent. In the {\it gauge fixed} theory we find that the seven kinematical
generators are manifestly independent of the interaction.

We shall work in the gauge $A^+=0$ and ignore all surface terms. Thus we are
working in the completely gauge fixed sector of the theory\cite{hk}. The
explicit form of the operator $J^3$ in this case is given in 
Ref. \cite{hk} which is manifestly free of interaction at the operator level.
The rotation operators are given in Sec. III.

At $x^+=0$, the operators $K^3$ and $E^i$ depend only on the density
$\Theta^{++}$. A straightforward calculation leads to 
\begin{eqnarray}
\Theta^{++} =  {\psi^+}^\dagger \st{\lh}{i\pp^+}\psi^+ + 
\partial^+ A^i \partial^+ A^i.
\end{eqnarray}
Then, longitudinal momentum operator,
\begin{eqnarray}
P^+ && = { 1 \over 2} \int dx^- d^2 x^\perp \Theta^{++} \nonumber \\
&& = { 1 \over 2} \int dx^- d^2 x^\perp \left [ {\psi^+}^\dagger 
\st{\lh}{i\pp^+} \psi^+ + \partial^+ A^j \partial^+ A^j \right ].
\end{eqnarray}
Generator of longitudinal scaling,
\begin{eqnarray}
K^3 && = - { 1 \over 4} \int dx^- d^2 x^\perp x ^- \Theta^{++}, \nonumber \\
&&=  - { 1 \over 4} \int dx^- d^2 x^\perp x ^- \left [
 {\psi^+}^\dagger \st{\lh}{i\pp^+}\psi^+ + \partial^+ A^j \partial^+ A^j
\right ].
\end{eqnarray}
Transverse boost generators,
\begin{eqnarray}
E^i && =  - { 1 \over 2} \int dx^- d^2 x^\perp x^i \Theta^{++}, \nonumber \\
&& = - { 1 \over 2} \int dx^- d^2 x^\perp x^i \left [  {\psi^+}^\dagger 
\st{\lh}{i\pp^+} \psi^+ + \partial^+ A^j \partial^+ A^j \right ].
\end{eqnarray}
The transverse momentum operator 
\begin{eqnarray}
P^i = { 1 \over 2} \int dx^- d^2 x^\perp \Theta^{+i}
\end{eqnarray}
which appears to have explicit interaction dependence.
Using the constraint equations for $\psi^-$ and $A^-$, we  have
\begin{eqnarray}
\Theta^{+i} && = \Theta^{+i}_F + \Theta^{+i}_G, \nonumber \\
\Theta^{+i}_F && = 2 {\psi^+}^\dagger i \partial^i \psi^+ + 2 g
{\psi^+}^\dagger A^i \psi^+, \\
\Theta^{+i}_G && = \partial^+ A^j \partial^i A^j - \partial^+ A^j \partial^j A^i
+ \partial^+ A^i \partial^j A^j - 2 g {\psi^+}^\dagger A^i \psi^+.
\end{eqnarray}
Thus 
\begin{eqnarray} 
P^i = { 1 \over 2} \int dx^- d^2 x^\perp \left [ 
 {\psi^+}^\dagger \st{\lh}{i\pp^i} \psi^+ +
  A^j \partial^+\partial^j A^i - A^i \partial^+\partial^j A^j 
- A^j\partial^+ \partial^i A^j  \right ].
\end{eqnarray}
Thus we indeed verify that all the kinematical operators are explicitly
independent of interactions. 

Lastly, the Hamiltonian operator can be written in the manifestly Hermitian
form as,
\begin{eqnarray}
P^- = {1 \over 2} \int dx^- d^2 x^\perp \Theta^{+-}
= {1\over 2}\int dx^-d^2x^\perp ( {\cal H}_0 + {\cal H}_{int}) 
\end{eqnarray}

where ${\cal H}_0$ is the free part given by,
\be
{\cal H}_0 =- A^j_a {(\pp^i)}^2 A^j_a + \xi^\dagger \Big [ {{-(\pp^\p)^2 +
m^2}\over {i\pp^+}}\Big ] \xi- \Big [ {{-(\pp^\p)^2 +
m^2}\over {i\pp^+}}\xi^\dagger\Big ] \xi .
\e   
The interaction terms are given by,
\be
{\cal H}_{int} = {\cal H}_{qqg} + {\cal H}_{ggg} + {\cal H}_{qqgg}+ 
{\cal H}_{qqqq} + {\cal H}_{gggg} ,
\e
where,
\be
 {\cal H}_{qqg} = -4g \xi^\dagger {1\over \pp^+} (\pp^\p.A^\p)\xi
 +  g{\pp^\p\over {\pp^+}} [ \xi^\dagger ({\tilde \s}^\p \cdot A^\p) ] {\tilde \s}^\p
\xi + g\xi^\dagger ({\tilde \s}^\p \cdot A^\p) {1\over {\pp^+}} ({\tilde
\s}^\p \cdot \pp^\p) \xi \nonumber\\ ~~~~~ + g({ \pp^\p \over {\pp^+}}
\xi^\dagger) {\tilde \s}^\p ( { \tilde \s}^\p \cdot A^\p) \xi + g\xi^\dagger
{1\over {\pp^+}} ({\tilde \s}^\p \cdot  \pp^\p) ( {\tilde \s}^\p \cdot
A^\p) \xi \nonumber\\ ~~~~~~ -mg {1\over {\pp^+}} [ \xi^\dagger ({\tilde
\s}^\p \cdot A^\p) ] \xi + m g\xi^\dagger ( {\tilde \s}^\p \cdot A^\p){1\over
{\pp^+}} \xi
\nonumber\\~~~~~~~~~~ + mg ( {1\over {\pp^+}} \xi^\dagger) ({\tilde \s}^\p
\cdot A^\p) \xi - mg \xi^\dagger {1\over {\pp^+}} [( {\tilde \s}^\p \cdot
A^\p) \xi] ,
\e
\be
{\cal H}_{ggg} = 2gf^{abc} \Big [ \pp^iA^j_aA^i_bA^j_c + (\pp^i
A^i_a){1\over {\pp^+}}(A^j_b\pp^+A^j_c)\Big ] ,
\e
\be
{\cal H}_{qqgg}&& = g^2 \Big [ \xi^\dagger ({\tilde \s}^\p.A^\p){1\over
{i\pp^+}} ({\tilde \s}^\p.A^\p) \xi - {1\over {i\pp^+}} (\xi^\dagger {\tilde
\s}^\p.A^\p) {\tilde \s}^\p.A^\p \xi
\nonumber\\&&~~~+ 4{1\over \pp^+} (f^{abc} A^i_b\pp^+A^i_c){1\over \pp^+}
(\xi^\dagger T^a \xi)\Big ] ,
\e 
\be
{\cal H}_{qqqq} = 4g^2 {1\over \pp^+} (\xi^\dagger T^a \xi){1\over \pp^+}
(\xi^\dagger T^a \xi) ,
\e
\be
{\cal H}_{gggg} = &&{g^2\over 2} f^{abc}f^{ade} \Big [ A^i_bA^j_cA^i_dA^j_e
\nonumber\\&&~~~~~~~~~+ 2{1\over \pp^+} ( A^i_b\pp^+A^i_c){1\over
\pp^+}(A^j_d \pp^+A^j_e)\Big ] .
\e
\section{Transverse Spin in free fermion field theory}
\subsection{Poincare Generators: Operator Forms}
The symmetric energy momentum tensor 
\begin{eqnarray} \Theta^{\mu \nu} =   \Big [{\overline \psi}
  \gamma^{\mu} {1 \over 4}
\st{\lh}{i\pp^\nu} \psi   +
 {\overline \psi} \gamma^{\nu}  {1 \over 4}\st{\lh}{i\pp^\mu} \psi \Big ] . \end{eqnarray}
\noindent The momentum operators are given by 
\begin{eqnarray}  P^{+}  && = 
 { 1 \over 2} \int dx^- d^2 x^\perp  {\overline \psi} \gamma^{+} 
{1 \over 2}\st{\lh}{i\pp^+}
  \psi \nonumber \\
&& = {1\over 2} \int dx^- d^2 x^\perp [\xi^{\dagger} i\partial^+  
 -(i\partial^+ \xi^{\dagger})]  \xi.
\end{eqnarray}
\begin{eqnarray} P^{i}  && = 
 { 1 \over 2} \int dx^- d^2 x^\perp
 \Bigg  [ {\overline \psi} \Big \{ \gamma^{+}{1 \over 4}
\st{\lh}{i\pp^i}  + 
 \gamma^{i} {1 \over 4}
\st{\lh}{i\pp^+}\Big \}  \psi \Bigg ]  \nonumber \\
&& = {1\over 2} \int dx^- d^2 x^\perp  [\xi^{\dagger} i \partial^{i}-(
i\partial^i \xi^\dagger )] \xi.
 \end{eqnarray}
 The Hamiltonian operator is  
\begin{eqnarray}  P^- && = 
{ 1 \over 2} \int  dx^- d^2 x^\perp   
{\overline \psi} \Big [ \gamma^-
  {1 \over 4} 
\st{\lh}{i\pp^+}
+ \gamma^+ { 1 \over 4}
\st{\lh}{i\pp^-}
\Big ] \psi \nonumber \\
&& = {1\over 2}\int dx^- d^2 x^\perp \Big [ \xi^{\dagger} {1 \over i \partial^{+}}[ m_F^2 -
(\partial^{\perp})^{2} ] - ({1 \over i \partial^{+}}[ m_F^2 -
(\partial^{\perp})^{2}]\xi^\dagger)\Big ]\xi
 . \end{eqnarray}
The longitudinal scaling operator (at $x^+=0$) is 
\begin{eqnarray} K^{3} && = - { 1 \over 2} \int dx^- d^2 x^\perp
 x^{-}  
\Bigg[{\overline\psi} \gamma^{+}{1 \over 4} \st{\lh}{i\pp^+} \psi \Bigg]  \cr
&& = - {i \over 4}  \int dx^- d^2 x^\perp x^{-} \big [\xi^{\dagger} \partial^{+}
\xi-(\partial^+ \xi^\dagger ) \xi \big ].  \end{eqnarray}
The transverse boost operators are
\begin{eqnarray} E^{i} && =  -{ 1 \over 4} \int dx^- d^2 x^\perp  x^{i}
 \Bigg [{\overline \psi} \gamma^+{1 \over 4} \st{\lh}{i\pp^+}\psi \Bigg ] \nonumber \\
&& = -  { 1 \over 4} \int dx^- d^2 x^\perp x^i\big [\xi^{\dagger}
i \partial^+ -(i\partial^+ \xi^\dagger)\big ]\xi
  . \end{eqnarray}
The generators of rotations are
\begin{eqnarray} J^{3} && ={ 1 \over 2 } \int dx^- d^2
x^\perp \Bigg \{
x^{1} \Bigg [ {\overline \psi} \Big \{ \gamma^{+}{ 1 \over 4} 
\st{\lh}{i\pp^2}+
 \gamma^{2}{ 1 \over 4} \st{\lh}{i\pp^+}\Big \} \psi \Bigg ] \nonumber \\
&&~~~~- x^{2} \Bigg [ {\overline \psi} \Big \{ \gamma^{+}{ 1 \over 4}
\st{\lh}{i\pp^1} 
 + 
\gamma^{1}{ 1 \over 4} \st{\lh}{i\pp^+}  \Big \}\psi \Bigg ] 
\Bigg \} \nonumber \\
&& = \int dx^- d^2 x^\perp \Big [\xi^{\dagger} 
\big [ {i\over 2} (x^1 {\overrightarrow \partial^2}- x^2 {\overrightarrow
\partial^1})\xi -\big [ {i\over 2} (x^1 {\overrightarrow \partial^2}- x^2 {\overrightarrow
\partial^1})\xi^\dagger\big ]\xi\nonumber\\&&~~~~~~~~~~~~~~~~~~~~~ 
 + \xi^\dagger{\sigma_{3} \over 2}\xi \big ].  \end{eqnarray}
 and 
\begin{eqnarray} F^{i}  && =  { 1 \over 2} \int dx^- d^2 x^\perp 
 \Bigg \{ 
x^{-} \Bigg [ {\overline \psi} \Big \{  \gamma^{+}{1\over 4} \st{\lh}{i\pp^i}
 + \gamma^{i}{1 \over 4} 
\st{\lh}{i\pp^+}
  \Big \} \psi \Bigg ] \nonumber \\
&&~~~- x^{i} \Bigg [{\overline \psi}\Big \{\gamma^{+}{1 \over 4} 
\st{\lh}{i\pp^-} 
+{1 \over 4} \gamma^{-} 
\st{\lh}{i\pp^+}
\Big \} \psi \Bigg ]
 \Bigg \} \nonumber \\
&& =   {i\over 2}  \int dx^- d^2 x^\perp \xi^{\dagger} \Bigg [ 
x^i (m^2 -(\partial^{\perp})^{2}){1 \over \partial^{+}} - x^{-}
{\partial \over \partial x^{i}} \nonumber \\
&& ~~~+ {1 \over \partial^{+}}
\Big \{ - { \partial \over \partial x^i} - i \epsilon^{ij} \sigma^3 {
\partial \over \partial x^j} + \epsilon^{ij} m\sigma^j \Big \} \Bigg ] \xi
\nonumber\\&&~~~- {i\over 2}  \int dx^- d^2 x^\perp  \Bigg [ 
\big [x^i (m^2 -(\partial^{\perp})^{2}){1 \over \partial^{+}} - x^{-}
{\partial \over \partial x^{i}} \nonumber \\
&& ~~~+ {1 \over \partial^{+}}
\Big \{ - { \partial \over \partial x^i} + i \epsilon^{ij} \sigma^3 {
\partial \over \partial x^j} + \epsilon^{ij} m\sigma^j \Big \}\xi^\dagger
\big ] \Bigg ] \xi
 . \end{eqnarray}
\subsection{Fock Representation}
Free spin-half field operator is
\begin{eqnarray} \xi(x) = \sum_{\lambda}\chi_\lambda 
 \int {dk^+ d^2 k^\perp \over 2 (2 \pi)^3 \sqrt {k^+}}  
[ b(k,\lambda)  e^{-ik.x}
+ d^{\dagger}(k,-\lambda)  e ^{ik.x}]  . \end{eqnarray}
In terms of Fock space operators
\begin{eqnarray} P^{+} = \int {dk^+ d^2 k^\perp \over 2 (2 \pi)^3 k^+} k^{+} \sum_{\lambda} \big 
[ b^{\dagger}(k, \lambda) b(k,\lambda)  +
d^{\dagger}(k,-\lambda) d(k,-\lambda) \big ].  \end{eqnarray}
\begin{eqnarray}
 P^{i} = \int {dk^+ d^2 k^\perp \over 2 (2 \pi)^3 k^+} k^{i} \sum_{\lambda} \big 
[ b^{\dagger}(k,\lambda) b(k,\lambda) + 
d^{\dagger}(k,-\lambda) d(k,-\lambda) \big ] . \end{eqnarray}
\begin{eqnarray}
 P^{-} = \int {dk^+ d^2 k^\perp \over 2 (2 \pi)^3 k^+} {m_F^2 + (k^{\perp})^2 \over k^{+}} \sum_{\lambda}
\big [ b^{\dagger}(k,\lambda) b(k,\lambda) + d^{\dagger}(k,-\lambda)
 d(k,-\lambda) \big ].  \end{eqnarray}
\begin{eqnarray} K^{3} &&=  - {i\over 2} \int {dk^+ d^2 k^\perp \over 2 (2 \pi)^3 k^+} k^{+} \sum_{\lambda}
\Big (\big [  {\partial b^{\dagger}(k,\lambda) \over 
\partial k^{+}} b(k,\lambda) +   {\partial
d^{\dagger}(k,-\lambda) \over \partial k^{+}}d(k,-\lambda) 
\big ]\nonumber\\&&~~~~~-\big [ b^{\dagger}(k,\lambda) {\partial b(k,\lambda)
\over {\partial k^+}} + 
d^{\dagger}(k,-\lambda){\partial d(k,-\lambda)\over {\partial k^+}} 
\big ] \Big ) . \end{eqnarray}
\begin{eqnarray} E^{i} &&= { i\over 2} \int {dk^+ d^2 k^\perp \over 2 (2 \pi)^3 k^+} \sum_{\lambda}
k^{+}\Big (\big [   {\partial b^{\dagger}(k,\lambda)
\over \partial k^{i}}b(k,\lambda)  +  
{\partial d^{\dagger}(k,-\lambda) \over \partial k^{i}} d(k,-\lambda)
 \big ]\nonumber\\&&~~~~~-\big [ b^{\dagger}(k,\lambda){\partial b(k,\lambda)
\over {\partial k^i}}  +  
d^{\dagger}(k,-\lambda){\partial d(k,-\lambda)\over {\partial k^i}}
 \big ] \Big ). 
\end{eqnarray}
\begin{eqnarray} J^{3} && = {i\over 2} \int {dk^+ d^2 k^\perp \over 2 (2 \pi)^3 k^+} \sum_{\lambda}
\big [ \Big( [ k^{1} {\partial \over \partial k^{2}}
- k^{2} {\partial \over \partial k^{1}} ] b^{\dagger}(k,\lambda)  \Big )
b(k,\lambda)-b^\dagger(k,\lambda) [ k^{1} {\partial \over \partial k^{2}}
- k^{2} {\partial \over \partial k^{1}} ] 
b(k,\lambda)
\nonumber\\&&~~~+ \Big ( \big [ k^{1} {\partial \over \partial k^{2}}
- k^{2} {\partial \over \partial k^{1}} ] d^{\dagger}(k,-\lambda)  \big ]
\Big )
 d(k,-\lambda)- d^\dagger(k,-\lambda)\big [ k^{1} {\partial \over \partial k^{2}}
- k^{2} {\partial \over \partial k^{1}} ]d(k,-\lambda) \big ]
 \nonumber \\
&&~~~~~~~~~~~~~~ + { 1 \over 2} \int {dk^+ d^2 k^\perp \over 2 (2 \pi)^3 k^+}
 \sum_{\lambda} \lambda \big [ b^{\dagger}(k,\lambda) b(k,\lambda) + 
d^{\dagger}(k,\lambda) d(k,\lambda) \big ] \end{eqnarray}
with $\lambda = \pm 1$. 
\begin{eqnarray} F^{i} && =   i \int {dk^+ d^2 k^\perp \over 2 (2 \pi)^3 k^+} k^i \sum_{\lambda}
\Big (\Big[ {\partial b^{\dagger}(k,\lambda)
\over \partial k^{+}} b(k,\lambda) +
{\partial d^{\dagger}(k,-\lambda)
\over \partial k^{+}} d(k,-\lambda) \Big ]\nonumber\\&&~~~~~~~~~~~~~-
\Big[ b^\dagger(k,\lambda){\partial b(k,\lambda)\over \partial k^{+}} +
d^\dagger(k, -\lambda){\partial d(k,-\lambda)
\over \partial k^{+}} \Big ]\Big )  \nonumber \\
&& + {i\over 2} \int {dk^+ d^2 k^\perp \over 2 (2 \pi)^3 k^+} 
 {m_F^2 + (k^{\perp})^2 \over k^{+}}\sum_{\lambda}
\Big ( \Big[ {\partial b^{\dagger}(k,\lambda)
\over \partial k^i} b(k,\lambda) +{\partial d^{\dagger}(k,-\lambda)
\over \partial k^i} d(k,-\lambda) \Big ]\nonumber\\&&~~~~~~~~~~~~~~~~
-\Big [ b^\dagger(k, \lambda){\partial b(k,\lambda)
\over \partial k^i} +
d^\dagger (k, -\lambda){\partial d(k, -\lambda) \over \partial k^i}
 \Big] \Big ) \nonumber \\&& - \int {dk^+ d^2 k^\perp 
\over 2 (2 \pi)^3 k^+}  { \epsilon^{ij} \over k^+}
 k^j \sum_{\lambda \lambda'}\sigma^{3}_{\lambda \lambda'}
\Big[ b^{\dagger}(k,\lambda) b(k,\lambda')
- d^{\dagger}(k,-\lambda') d(k,-\lambda)\Big ] \nonumber \\
&& - \int {dk^+ d^2 k^\perp \over 2 (2 \pi)^3 k^+}  
{ \epsilon^{ij} \over k^+}
 m_F \sum_{\lambda \lambda'}\sigma^{j}_{\lambda \lambda'}
\Big[ b^{\dagger}(k,\lambda) b(k,\lambda')
+ d^{\dagger}(k,-\lambda) d(k,-\lambda')\Big ]. 
\end{eqnarray}
\subsection{Transverse Spin of a Single Fermion}
For a single fermion of mass $m$ and momenta ($k^+,k^\perp$), we have,
\begin{eqnarray}
P^+ \mid k \lambda \rangle &&=k^+ \mid k \lambda \rangle,
~P^1\mid k \lambda \rangle=k^1\mid k \lambda \rangle,
 ~ P^2\mid k \lambda \rangle=k^2\mid k \lambda \rangle, \nonumber \\
 P^- \mid k \lambda \rangle &&= {(k^\perp)^2 + m^2 \over k^+}\mid k \lambda \rangle,
~{\cal J}^3\mid k \lambda \rangle={1 \over 2} \lambda \mid k \lambda \rangle, 
\nonumber \\
~K^3 \mid k \lambda \rangle &&= -i k^+ { \partial \over \partial k^+}
\mid k \lambda \rangle, ~ E^1
\mid k \lambda \rangle = i k^+ { \partial \over
\partial k^1}
\mid k \lambda \rangle, ~E^2 
\mid k \lambda \rangle= i k^+ { \partial \over \partial k^2}
\mid k \lambda \rangle, \nonumber \\
F^1 \mid k \lambda \rangle &&= \left (2 i k^1 {\partial \over \partial k^+} + i {(k^\perp)^2 + m^2 \over k^+}
{\partial \over \partial k^1} - {k^2 \over k^+} \lambda \right ) \mid k
\lambda \rangle  - { m
\over k^+} \sum_{\lambda'}\sigma^2_{\lambda' \lambda} 
\mid k \lambda' \rangle, \nonumber \\
F^2 \mid k \lambda \rangle && = \left (2 i k^2 {\partial \over \partial k^+} + i {(k^\perp)^2 + m^2 \over k^+}
{\partial \over \partial k^2} + {k^1 \over k^+} \lambda \right )
\mid k \lambda \rangle
  + { m \over k^+} \sum_{\lambda'} \sigma^1_{\lambda' \lambda} 
\mid k \lambda' \rangle.
\end{eqnarray}
We arrive at 
\begin{eqnarray}
m{\cal J}^1
\mid k \lambda \rangle &&= \left ({ 1 \over 2} F^2 P^+ +K^3P^2 
 - { 1 \over 2} E^2 P^- - P^1
{\cal J}^3 \right ) 
\mid k \lambda \rangle\nonumber \\
&& = m \sum_{\lambda'}{\sigma^1_{\lambda' \lambda} \over 2}
\mid k \lambda' \rangle,
\end{eqnarray}
\begin{eqnarray}
m {\cal J}^2 
\mid k \lambda \rangle&& = \left (- { 1 \over 2} F^1 P^+- K^3 P^1  + { 1 \over 2} E^1 P^- -
P^2 {\cal J}^3 \right )
\mid k \lambda \rangle\nonumber \\
&& = m \sum_{\lambda'}{\sigma^2_{\lambda' \lambda} \over 2}
\mid k \lambda' \rangle.
\end{eqnarray} 

\section{Transverse spin in free massless spin one field theory}
\subsection{Poincare Generators: Operator Forms}
The symmetric gauge invariant energy momentum tensor
\begin{eqnarray} \Theta^{\mu \nu} = 
F^{\lambda \mu} F^{\nu}_{\lambda} - g^{\mu \nu} {\cal L} . \end{eqnarray}
where the Lagrangian density  
\begin{eqnarray} {\cal L} = -{1 \over 4} F^{\mu \nu} F_{\mu \nu}
\end{eqnarray}
with
\begin{eqnarray} F^{\mu \nu} = \partial^{\nu} A^{\mu} - \partial^{\mu} A^{\nu}
. \end{eqnarray}

We choose $A^{+}=0$ gauge. Only the transverse fields $A^{i}$ are dynamical
variables.  
The momentum operators are given by
\begin{eqnarray} P^{+} = { 1 \over 2}\int dx^- d^2 x^\perp  \partial^{+} A^{j}
\partial^{+} A^{j}   , \end{eqnarray}
\begin{eqnarray} P^i = { 1 \over 2} \int dx^- d^2 x^\perp \Big (
A^{j}\partial^+\partial^j A^{i} - A^{i}\partial^+ \partial^j A^{j}
-A^{j}\partial^+\partial^i A^{j}\Big ). \end{eqnarray}
The Hamiltonian operator is 
\begin{eqnarray} P^- &&= 
{1 \over 2} \int dx^- d^2 x^\perp \Big 
[ {1 \over 4} (\partial^+ A^-)^2 + { 1 \over 2} F^{ij} F_{ij} \Big ]
\nonumber \\
&& = {1 \over 2}\int dx^- d^2 x^\perp  \partial^{i} A^{j} \partial^{i} A^{j}  
  = -{1\over 2}\int dx^- d^2x^\perp A^j {(\partial^i)}^2 A^j
\end{eqnarray}
The longitudinal scale generator (at $x^+=0$) is  
\begin{eqnarray}  K^3 = -{1 \over 2} \int dx^- d^2 x^\perp x^{-}  \partial^{+} A^{j} \partial^{+} A^{j} 
 . \end{eqnarray}
The transverse boost generators are
\begin{eqnarray} E^{i} = - {1 \over 2}\int dx^- d^2 x^\perp x^{i}  \partial^{+} A^{j} \partial^{+} A^{j} 
. \end{eqnarray}
The generators 
 of rotations are
\begin{eqnarray} J^3 &&= {1 \over 2}\int dx^- d^2 x^\perp \Big ( 
x^1 [ \partial^+ A^2  \partial^i A^i + \partial^+ A^1
( \partial^2 A^1 - \partial^1 A^2)] \nonumber \\
&& - x^2 [ \partial^+ A^1  \partial^i A^i + \partial^+ A^2
(- \partial^2 A^1 + \partial^1 A^2)] \Big)\nonumber \\
&& = { 1 \over 2} \int dx^- d^2 x^\perp \left (x^1 
[ \partial^+ A^1 \partial^2 A^1 + \partial^+ A^2 \partial^2 A^2]
-x^2 [ \partial^+ A^1 \partial^1 A^1 + \partial^+ A^2 \partial^1 A^2] \right )
\nonumber \\
&& ~~~ + { 1 \over 2} \int dx^- d^2 x^\perp 
[ A^1 \partial^+ A^2 - A^2 \partial^+ A^1 ]. 
   \end{eqnarray}
and 
\begin{eqnarray} F^i && = {1 \over 2}\int dx^- d^2 x^\perp \Big (
x^- \Big (
A^{ja}\partial^+\partial^j A^{i} - A^{i}\partial^+  \partial^j A^{j}
-A^{j}\partial^+\partial^i A^{j}\Big ) \nonumber \\
&& - x^i [  A^k (\partial^j)^2 A^k] \Big ) 
- 2\int dx^- d^2 x^\perp A^i \eta^{ij} \partial^j A^j, ~~{\rm
no~summation~over}~i,j ~
, \end{eqnarray}
with $ \eta^{12}=\eta^{21}=1$, $\eta^{11}=\eta^{22}=0$.
\subsection{Fock Representation}
The dynamical components of the free massless spin field operator 
in $A^{+}=0$ gauge are
\begin{eqnarray}  A^{i}(x) = \sum_{\lambda=1}^{2} \int {dk^+ d^2 k^\perp \over 2 (2 \pi)^3 k^+} \delta^{i \lambda}
[ a(k,\lambda)e^{-ik.x} + a^{\dagger}(k,\lambda) e^{ik.x} ]. \end{eqnarray}
In terms of Fock space operators, we have,
\begin{eqnarray} P^{+} =  \int {dk^+ d^2 k^\perp \over 2 (2 \pi)^3 k^+} k^{+} \sum_{\lambda}  
a^{\dagger}(k,\lambda) a (k,\lambda).  \end{eqnarray}
\begin{eqnarray} P^{i} =  \int {dk^+ d^2 k^\perp \over 2 (2 \pi)^3 k^+} k^{i} \sum_{\lambda} 
a^{\dagger}(k,\lambda) a(k,\lambda)  . \end{eqnarray}

 \begin{eqnarray} H =  \int {dk^+ d^2 k^\perp \over 2 (2 \pi)^3 k^+} 
 {{k^{\perp}}^{2} \over k^{+}}\sum_{\lambda} 
a^{\dagger}(k,\lambda) a(k,\lambda)  . \end{eqnarray}
\begin{eqnarray} K^3 = - { i\over 2}\int {dk^+ d^2 k^\perp \over 2 (2 \pi)^3 k^+} 
 k^{+}\sum_{\lambda} \Big [
\big ( {\partial a^{\dagger}(k,\lambda) \over \partial
k^{+}}\big )  a(k,\lambda)- a^\dagger ( k, \lambda){\partial a(k,\lambda)
 \over \partial k^{+}}\big )\Big ]  . \end{eqnarray}
\begin{eqnarray} E^{i} =  {i\over 2}  \int {dk^+ d^2 k^\perp \over 2 (2 \pi)^3 k^+} 
k^{+}\sum_{\lambda} \big [
 \big ({\partial a^{\dagger}(k,\lambda) \over \partial k^{i}} \big )
a(k,\lambda)- a^\dagger ( k, \lambda){\partial a(k,\lambda)
 \over \partial k^{i}}\big ) \big ] 
 . \end{eqnarray}
\begin{eqnarray} J^3 && = 
{i\over 2}  \int {dk^+ d^2 k^\perp \over 2 (2 \pi)^3 k^+} \sum_{\lambda}\big
[ \Big ( 
(k^1 {\partial \over \partial k^2} - k^2 {\partial \over \partial k^1}
) a^{\dagger}(k,\lambda) \Big ) a(k,\lambda)-a^\dagger (k, \lambda)(k^1 {
\partial \over \partial k^2} - k^2 {\partial \over \partial k^1}
) a(k,\lambda)\big ] \nonumber \\
&&~~~~~~~~~~~~~~~~~~~~~~~ + i \int {dk^+ d^2 k^\perp
 \over 2 (2 \pi)^3 k^+}  \Big (
a^{\dagger}(k,2) a(k,1) - a^{\dagger}(k,1) a(k,2) \Big )  . \end{eqnarray} 
Introduce creation and annihilation operators 
\begin{eqnarray} a(k, \uparrow) = { - 1 \over \sqrt{2}} [ a(k,1) - i a(k,2)] ,
 a(k, \downarrow) = {  1 \over \sqrt{2}} [ a(k,1) + i a(k,2)]
.\end{eqnarray}
Then
\begin{eqnarray} J^3 && = {i\over 2}
 \int {dk^+ d^2 k^\perp \over 2 (2 \pi)^3 k^+} \sum_{\lambda}\big [ \Big ( 
(k^1 {\partial \over \partial k^2} - k^2 {\partial \over \partial k^1}
) a^{\dagger}(k,\lambda) \Big ) a(k, \lambda)- a^\dagger (k, \lambda)(k^1 {
\partial \over \partial k^2} - k^2 {\partial \over \partial k^1}
) a(k,\lambda)\big ] \nonumber \\
&&~~~~~~~~~~~~~~~~~~~~ +  \int {dk^+ d^2 k^\perp \over 2 (2 \pi)^3 k^+}
 \sum_{\lambda} \lambda a^{\dagger}(k,\lambda) a(k,\lambda)   . \end{eqnarray} 

where $\lambda$ now denotes circular polarization.
\begin{eqnarray}  F^i = &&
 i  \int {dk^+ d^2 k^\perp \over 2 (2 \pi)^3 k^+} k^i
\sum_{\lambda} \big ({\partial a^{\dagger}(k,\lambda) \over
\partial k^{+}} a(k,\lambda) - a^\dagger ( k, \lambda){\partial a(k,\lambda)
 \over \partial k^{+}}\big ) \nonumber \\ 
&& + {i\over 2}  \int {dk^+ d^2 k^\perp \over 2 (2 \pi)^3 k^+}
 {(k^{\perp})^2 \over k^+}\sum_{\lambda} 
\big ( {\partial a^{\dagger}(k,\lambda) \over \partial k^i}a( k, \lambda) -a^\dagger 
( k, \lambda){\partial a(k,\lambda) \over \partial k^{i}}\big )
 \nonumber \\
&& - 2  \epsilon^{ij} \int {dk^+ d^2 k^\perp \over 2 (2 \pi)^3 k^+}
{k^j \over k^+} \sum_\lambda \lambda a^\dagger(k,\lambda) a(k,\lambda).  
\end{eqnarray}
\subsection{Transverse Spin}
Using the explicit form of the operators, we get for a state of
momentum $k(k^+, k^\perp)$ and helicity $\lambda$, 
\begin{eqnarray}
{\cal J}^3 \mid k \lambda \rangle ={W^+ \over P^+} \mid k \lambda \rangle &&= \lambda \mid k \lambda \rangle,
\nonumber \\
W^1 \mid k \lambda \rangle &&= k^1 \lambda \mid k \lambda \rangle, \nonumber \\
W^2 \mid k \lambda \rangle &&= k^2 \lambda \mid k \lambda \rangle, \nonumber \\
W^- \mid k \lambda \rangle && = {(k^\perp)^2 \over k^+} \lambda \mid k
\lambda \rangle.
\end{eqnarray}
\begin{eqnarray}
{\cal J}^i \mid k \lambda \rangle =0.
\end{eqnarray}


\end{document}